\documentclass[aps,prl,reprint,
							floatfix,
							superscriptaddress,
							footinbib,
							floatfix,
							longbibliography]{revtex4-1}

\usepackage{booktabs}

\usepackage{amsmath}
\usepackage{amssymb}
\usepackage[normalem]{ulem}

\usepackage{natbib}
\usepackage[T1]{fontenc}
\usepackage[latin1]{inputenc}

\usepackage[dvipsnames]{xcolor}

\usepackage{hyperref}
\hypersetup{colorlinks=true,citecolor=RoyalBlue,urlcolor=Maroon}

\usepackage[squaren]{SIunits}

\usepackage[pdftex]{graphicx}
\usepackage{grffile} % allows dots in graphic file names
\usepackage{subfigure}

\usepackage{slashed}
\usepackage{todonotes}

\let\vec\boldsymbol

\newcommand{\ud}{\mathrm{d}}

\begin{document}

\title{Ultrafast Polarization of an Electron Beam in an Intense Bi-chromatic Laser Field}

\author{Daniel Seipt}
\email{dseipt@umich.edu}
\affiliation{The Gérard Mourou Center for Ultrafast Optical Science, University of Michigan, Ann Arbor, Michigan 48109, USA}

\author{Dario Del Sorbo}
\affiliation{High Energy Density Science Division, SLAC National Accelerator Laboratory, Menlo Park, CA 94025, USA}

\author{Christopher P. Ridgers}
\affiliation{York Plasma Institute, Department of Physics, University of York, York YO10 5DD, United Kingdom}

\author{Alec G. R. Thomas}
\affiliation{Center for Ultrafast Optical Science, University of Michigan, Ann Arbor, Michigan 48109, USA}

\begin{abstract}
Here, we demonstrate the radiative polarization of high-energy electron beams in collisions with ultrashort pulsed bi-chromatic laser fields.  Employing a Boltzmann kinetic approach for the electron distribution 
allows us to simulate the beam polarization over a wide range of parameters and determine the optimum conditions for maximum radiative polarization.
Those results are contrasted with a Monte-Carlo algorithm where
photon emission and associated spin effects are
treated fully quantum mechanically using spin-dependent photon emission rates.
The latter method includes realistic focusing laser fields, which allows us to simulate a near-term experimentally feasible scenario of a 8~GeV electron beam scattering from a 1 PW laser pulse and provide a measurement that would verify the ultrafast radiative polarization in high-intensity laser pulses that we predict.
Aspects of spin dependent radiation reaction are also discussed, with spin polarization leading to a measurable (5\%) splitting of the energies of spin-up and spin-down electrons.

\end{abstract}

\date{\today}

\maketitle

One of the driving forces in the development of petawatt class laser systems \cite{ELI-NP,Sung:OptLett2017,SEL100PW} is related to novel laser-plasma based accelerator concepts. Recent laser wakefield acceleration experiments have demonstrated acceleration of electrons up to 8~GeV  energy in a single stage \cite{Bella8GeV} and may enable future novel TeV electron-positron colliders
for high-energy physics \cite{LPWA-TEV,Schroeder_NIMPRA_2016}. 
Spin-polarized beams are crucial for high-energy collider applications, for instance in order to suppress the standard-model background in searches for new physics beyond the standard model \cite{Moortgat-Pick:PhysRep2008,Vauth:JMPConfSer2016}. Studying the dynamics of spin-polarized electrons in a plasma wakefield is therefore important \cite{Vieira:PRSTAB2011,Wen2018,Wu2019}. Spin forces have been suggested to be important in both astrophysical systems and high intensity laser-plasma interactions \cite{Mahajan_MNRAS_2015}. With increasing interest in high intensity laser-plasma interactions, it is essential to understand lepton spin effects on the overall plasma dynamics through radiation reaction---because of the spin-dependence in the photon emission rates---and electron-positron pair generation \cite{Ritus:JSLR1985}. There has been notable recent interest in spin polarization of leptons in high-intensity laser interactions \cite{Li:PRL2019,Chen2019,Wan2019}.

\begin{figure}[t]
    \centering
    \includegraphics[width=\columnwidth]{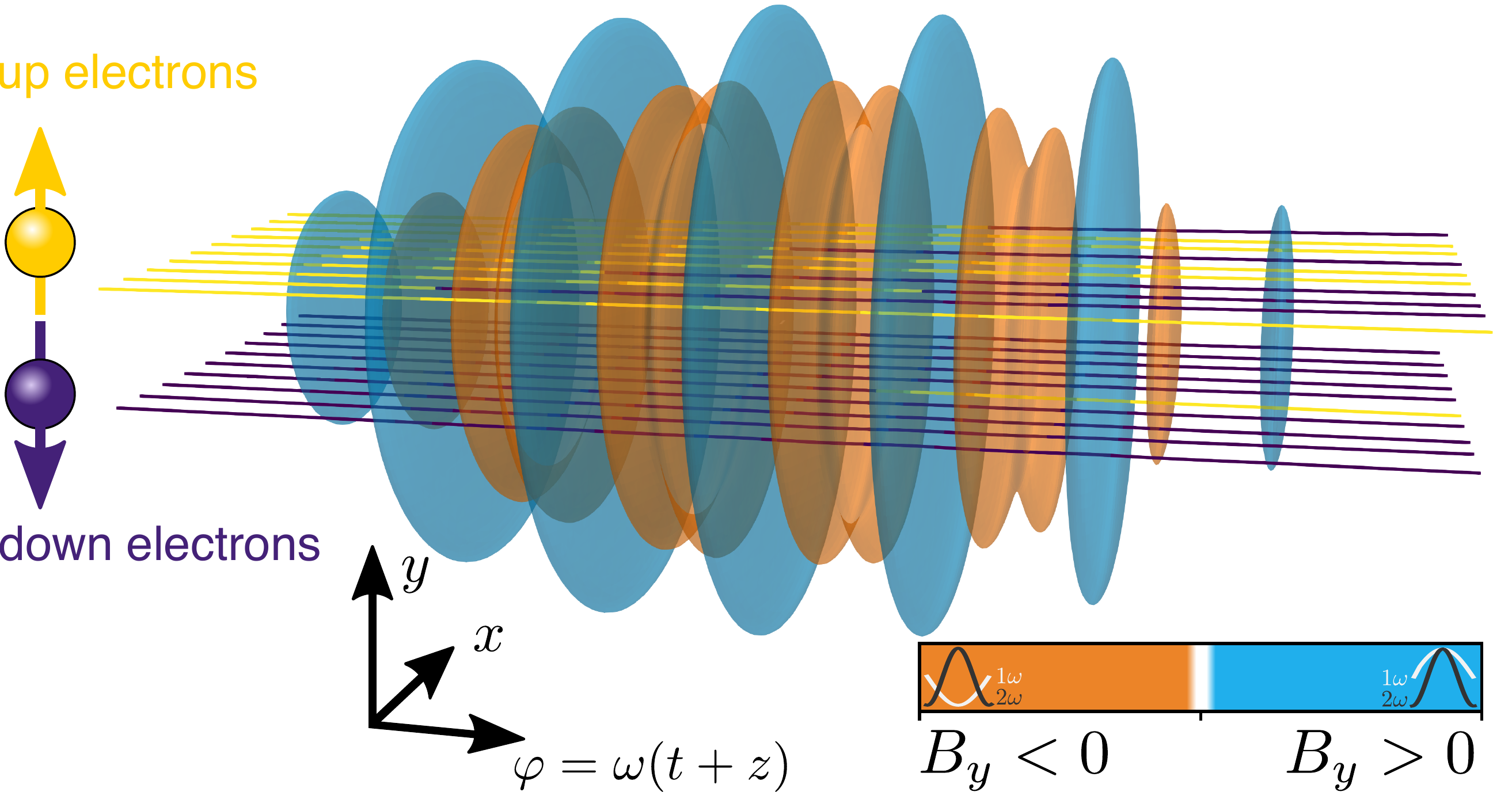}
\caption{Electrons propagating through a bi-chromatic laser pulse perform spin-flips dominantly in certain phases $\varphi$ of the field:
Electrons initially polarized along the $+y$ direction (yellow trajectories) flip their spin to down (trajectory colored purple) dominantly when $B_y>0$, and this is where $1\omega$ and $2\omega$ add constructively (blue contours). The opposite spin-flip dominantly happens when $B_y<0$ where the $1\omega$ and $2\omega$ components of the laser are out of phase  (orange contours).
Therefore $\chi$ is larger for $\uparrow \downarrow$-flips and many more of those flips happen than the $\downarrow \uparrow$-flips,
causing a polarization of the beam.}
    \label{fig:fig1}
\end{figure}

It is known that initially unpolarized lepton beams radiatively polarize in storage rings due to asymmetries in the rates for synchrotron emission; the so-called Sokolov-Ternov effect
\cite{book:Sokolov,Ternov:PhysUspekh1995,Mane:RPP2005}. As a result, their projected spins end up being predominantly anti-aligned with the magnetic field direction. 
For most machines this is a slow process, with a timescale of minutes or hours, scaling as $T \propto \gamma/\chi^3$ \cite{Ternov:PhysUspekh1995,Mane:RPP2005}, 
where $\chi = \sqrt{e^2 \, p.F^2.p}/m^3 = \gamma |e| B/m^2 \ll 1$ 
is the quantum nonlinearity parameter, with the field strength tensor $F$ and electron 
mass $m$, charge $e$ and four-momentum $p^\mu$ with Lorentz factor $\gamma = p^0/m$
\footnote{We use natural Heaviside-Lorentz units with $\hbar=c=\epsilon_0=1$, with fine structure constant $\alpha = e^2/4\pi$.
Lorentz indices are omitted whenever possible, e.g.~$u^\mu\to u$ and $f^{\mu\nu}\to f$, and we denote scalar products and contractions of tensor indices as $f^{\mu\nu} u_\nu \to f.u$.}.

We previously demonstrated that a similar spin-polarization can occur for electrons circulating at the magnetic-nodes of two colliding intense laser pulses \cite{DelSorbo:PRA2017}, with polarization timescales of few femtoseconds for laser intensities exceeding
$\unit{5 \times  10^{22}}{\watt/\centi\metre^2}$. However, these orbits are unstable in general \cite{DelSorbo:PPCF2018}. By first accelerating electrons to high energy and then colliding them with a laser pulse, it is possible to achieve $\chi\gtrsim1$ 
with current PW class high intensity lasers operating at intensities
$\sim \unit{10^{21}}{ \watt\per \centi\metre^2 }$ and study radiative spin polarization 
in the strongly quantum regime. 
This laser-electron-beam collider setup was used in seminal SLAC E-144 experiments demonstrating nonlinear Compton scattering \cite{Bula_PRL_1996} and electron-positron pair production \cite{Burke_PRL_1997} and has recently been used to observe quantum radiation reaction \cite{Cole:PRX2018,Poder:PRX2018}. 
In Ref.~\cite{DelSorbo:PPCF2018}, we discussed a polarization dependence of the radiation reaction force, related to the fact that spin-down electrons radiate more power than spin-up electrons.

In this Letter, we propose using bi-chromatic laser fields to polarize an electron beam and predict a measurable modification of the resulting quantum radiation reaction, Fig.~\ref{fig:fig1}. 
Spin-dependent radiation-reaction effects are described in our model using
(i) a kinetic approach where we solve a Boltzmann equation for distribution functions of spin polarized electrons and (ii) a quasi-classical particle tracking approach where electrons are pushed classically between photon emissions, and the emissions are treated fully quantum mechanically using a Monte Carlo algorithm employing spin-dependent photon emission rates. We determine optimum parameters for maximum radiative polarization, and discuss spin dependent radiation reaction leading to a measurable splitting of the mean electron energies.

For multi-GeV electrons colliding with a high-power laser pulse the emission
of gamma photons can be described as nonlinear Compton scattering using strong-field QED in the Furry picture \cite{Ritus:JSLR1985,DiPiazza:RevModPhys2012}. 
The quantum-radiation dominated regime is reached for normalized laser amplitude $a_0=eE/m\omega \gg 1$ and $\chi\sim 1$. The short photon formation-length implies that quantum emissions can be described as incoherent events in a locally constant field approximation (LCFA).
Hence,  LCFA photon emission rates are customarily employed in simulations         
\cite{Nerush:PRL2011,
        Bulanov:PRA2013,
        Ridgers:JCompPhys2014,
        Vranic:NJP2016,
        Harvey:PRA2015,
        Gonoskov:PRE2015,
        King:PRA2013,King:PRA2015,
        Blackburn:PhysPlas2018},
but see also \cite{DiPiazza:PRA2018,
                    Ilderton:PRA2019,
                    Ilderton:PRD2019,
                    Podszus:PRD2019}.

In a laser pulse, the magnetic field oscillates, and so in contrast with the case of a static magnetic field, the polarization built-up in one half cycle is mostly lost during the following half cycle. In Ref.~\cite{Seipt:PRA2018} we investigated an laser-electron-beam collider scenario, where an asymmetry between the two half-cycles is introduced by using ultra-short sub-cycle laser pulses. In that case, however, only a very small degree of polarization could be achieved due to the short interaction duration. A similar effect had been predicted due to electron self-interaction \cite{Meuren:PRL2011}. In Ref.~\cite{Seipt:PRA2018} we also showed the agreement between the LCFA and full QED S-matrix calculations for spin-polarization effects. In the scheme proposed here, the addition of a $2\omega$ component to the laser breaks the symmetry and allows for net radiative polarization of the beam without the necessity of sub-cycle pulses.

For electrons in an external field the canonical choice for the spin quantization axis 
is the space-like axial 4-vector-field
$\zeta = \zeta(x,p) =  \tilde f.p/(p.\tilde f^2.p)^{1/2} = 
    (\vec u\cdot \vec \zeta^\mathrm{RF} ,  
    \vec \zeta^\mathrm{RF} 
        + \vec u ( \vec u\cdot \vec \zeta^\mathrm{RF})/(\gamma(1+\gamma) ) $ 
        \cite{book:Landau4,Mane:RPP2005},
where $\vec \zeta^\mathrm{RF}$ points along the magnetic field in the rest frame of the electron, and with $\tilde f$ the dual of the normalized field strength tensor $f=eF/m$.
During the emission of a photon the electron spin will be projected onto a quantum eigenstate, with spin projection eigenvalue $\pm1$ along the local value of $\zeta$ \cite{Mane:RPP2005,Li:PRL2019}. Between emissions, the electrons are assumed to follow classical trajectories, with four-velocity $u = (\gamma,\vec u) = p/m$ governed by the Lorentz force equation, $\ud u / \ud \tau = f.u$
(spin-gradient (Stern-Gerlach) forces are negligible here \cite{Tamburini:NJP2010,DelSorbo:PPCF2018}), 
and the precession of the spin vector expectation value $S$ governed by the Thomas-Bargman-Michel-Telegdi (T-BMT) equation \cite{Bargmann:PRL1959}, $\ud S / \ud \tau = \frac{g_e}{2} f.S + \frac{g_e-2}{2} \,(S.f.u) \, u $,
with gyro\-magnetic ratio $g_e$, and with initial conditions $S=\pm \zeta$.

While the spin described above is a property of an individual electron, the polarization is a property of the whole beam \cite{Mane:RPP2005}, which can be 
defined as the ensemble average
$\mathbb S = \langle  S_y^\mathrm{RF} \rangle 
%= \langle S^y - S^0 u^y/(\gamma+1) \rangle 
\approx \langle  S_y \rangle $
for the scattering scenario discussed in this letter.

%\paragraph{Kinetic Equations---}%
Assuming the bi-chromatic laser field is a plane wave propagating along the negative $z$-axis and polarized along the $x$-axis, then the magnetic field, $\vec \zeta^\mathrm{RF}$ and $\zeta$
all have non-vanishing $y$-components only. We can therefore define two distributions, $n^s$, $s=\pm1$, of electrons with spins projected on the $y$-axis. 
The evolution of the distributions $n^s$ and transitions between spin states, 
$n^{+1}\leftrightarrow n^{-1}$, may be described by Boltzmann equations with transitions due to radiation emission described by means of a collision-like integro-differential operator \cite{Neitz:PRL2013,Ridgers:JPP2017}.

For electrons colliding with a plane-wave laser pulse, the light-front momentum 
$p^+ = p^0 + p^z$ is conserved under the classical Lorentz equation, with their dynamics solely determined by quantum radiation reaction effects \cite{Harvey:PRD2011b}. 
We can thus write the kinetic equations for the distribution functions, $ n^s(p^+,\varphi)$,
in the 1D approximation as
\begin{multline} \label{eq:KE}
\frac{\partial n^s}{\partial \varphi }  =
					 	\sum_{s'}\left[
                        \int_{0}^\infty \! \ud k^+ \:
					  n^{s'}(q^+)
					  \frac{\ud \mathbb R^{s'_\zeta s_\zeta} }{\ud k^+}(q^+)-  n^s \mathbb  R^{s_\zeta s'_\zeta } \right] \,,
\end{multline}
with laser phase $\varphi=\omega(t+z)$ 
%as the natural evolution parameter \cite{Neville:PRD1971}%
and $q^+ = p^++k^+$.
The spin-dependent probability per unit laser-phase for photon emission, i.e. the LCFA photon emission rate, appearing in the collision operator in \eqref{eq:KE} is given by \cite{Seipt:PRA2018}
\begin{multline} \label{eq:rates}
    \frac{\ud \mathbb R^{s_\zeta s_{\zeta'} }}{\ud k^+} 
    = 
    - \frac{\alpha m^2 }{ \omega (p^+)^2 } 
    \left[ (1 + s_\zeta s_{\zeta'}  ) \mathrm{Ai}_1(z) \vphantom{\frac{1}{2}}
    \right.  \\
    + ( g +  s_\zeta s_{\zeta'} ) \frac{2 \mathrm{Ai}'(z)}{z} 
    \left.
    + \left( s_\zeta t + s_{\zeta'} \frac{t}{1-t} \right) \frac{\mathrm{Ai}(z)}{\sqrt{z}} \right] \,.
\end{multline}
Here $s_\zeta$ ($s_{\zeta'}$) denotes the value of the electron spin projection
along the magnetic field in the instantaneous electron rest frame before (after) the photon emission, e.g.~$\zeta = \tilde f.p/(p.\tilde f^2.p)^{1/2}$ with 
$p$ as the instantaneous electron 4-momentum prior to photon emission,
and $\chi_p = (p.f^2.p)^{1/2}/m^4$. 
In Eq.~\eqref{eq:KE} $s_\zeta^{(\prime)} = s^{(\prime)} \,  {\rm sgn}(B_y)$.

The argument of the Airy functions, its derivative and integral $\mathrm{Ai}_1(z) = \int_z^\infty \ud x \, \mathrm{Ai}(x)$ is $z = [ t/(\chi_p (1-t)) ]^{2/3}$,
with the normalized light-front momentum of the emitted photon $t = k^+/p^+ \in (0,1)$, and $g=1+t^2/(2-2t)$.
For typical high-energy electron-beam laser collisions
$\zeta$ agrees with the local 2nd binormal Frenet-Serret vector up to terms
$(\mathrm{tr} \tilde f.f )^2/ m^4 \chi^2 \lll 1$
ensuring that $s_\zeta$ is a constant of motion in a constant crossed field, which is required for the applicability of the spin-dependent LCFA rates \eqref{eq:rates} in simulation codes \cite{Honig:JMathPhys1974,Seipt:PRA2018,Seipt:2019}.

The 1D approximation holds when $p_\perp \ll p^+$,  the
transverse quiver amplitude of electrons in the field $  a_0/\gamma\omega$ is much smaller than the laser spot size $w_0$ \cite{Neitz:PRL2013,Bulanov:PRA2013}, and that $\zeta$ is not precessing under the T-BMT equation.
These kinetic equations were validated for the case of a uniform static magnetic field by comparison with Sokolov-Ternov rate equations \cite{book:Synge,Ternov:PhysUspekh1995}.

\begin{figure}
    \centering
    \includegraphics[width=\columnwidth]{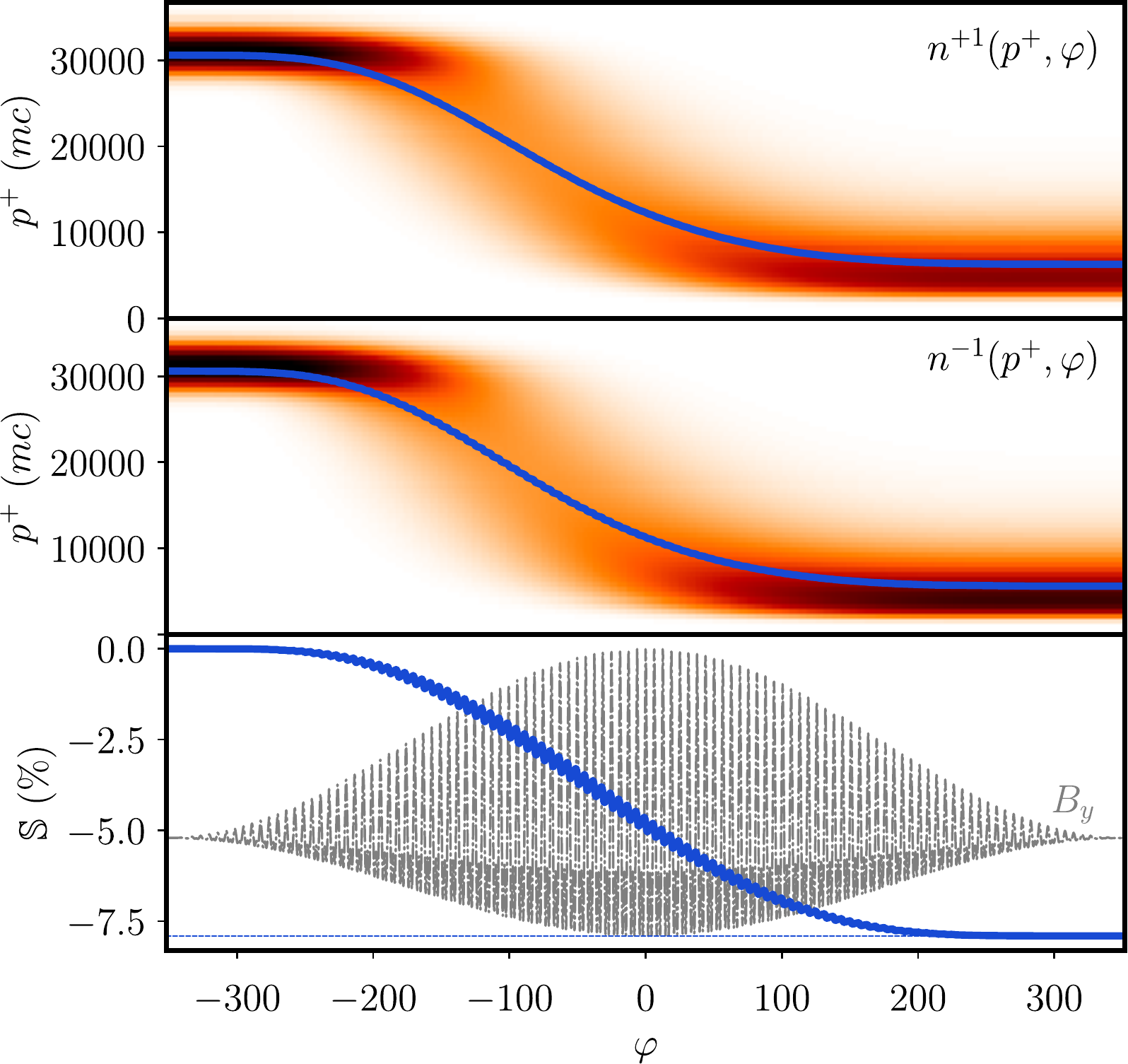}
    \caption{
    Kinetic equation solution for a 7.8 GeV electron beam \cite{Bella8GeV}
    colliding with a laser with $a_0=10.8$, $c_2=0.7$, $109$ fs FWHM duration, $\chi_0=1$. Evolution of $n^{+1}$, $n^{-1}$ and $\mathbb S$
    as function of laser phase from top to bottom.
    $n^{+1}$ means the spin is aligned along the $+y$ axis. The shape of the magnetic field $B_y$ is shown as thin grey curve in the lower plot.}
    \label{fig:BELLA-KE8}
\end{figure}

\begin{figure}
    \centering
    \includegraphics[width=\columnwidth]{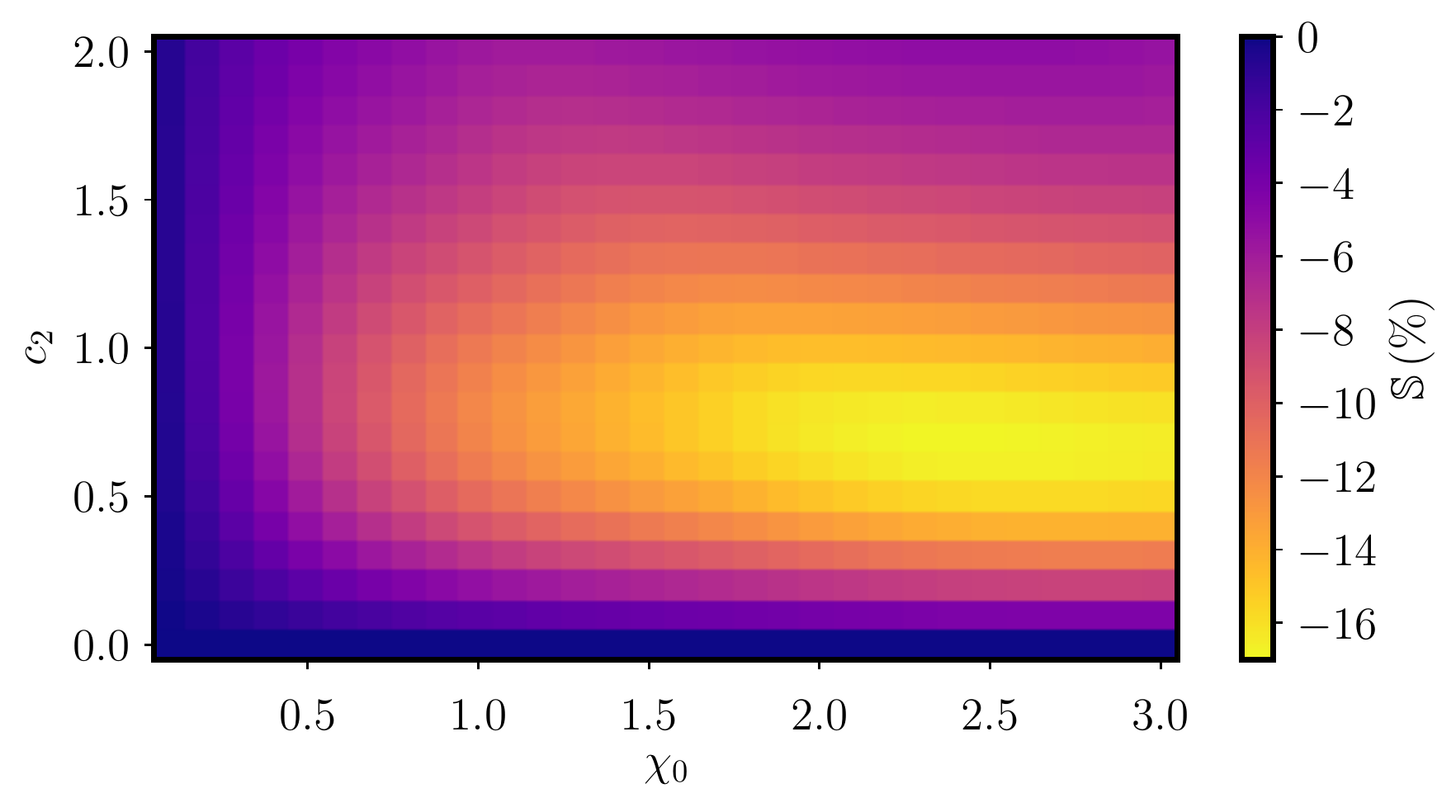}
    \caption{Achievable degree of electron polarization as a function of $\chi_0$ and bi-chromaticity parameter $c_2$.
    Calculations have been performed for $\unit{5}{\giga\electronvolt}$ electrons colliding with a
    $\unit{161}{\femto\second}$ laser pulse, i.e.~$a_0(\chi_0=1)=16.5$.
    }
    \label{fig:survey}
\end{figure}

The bi-chromatic field used to solve Eq.~\eqref{eq:KE} is characterized by the function
$\psi(\varphi)=  [\cos\varphi + c_2 \cos ( 2\varphi +\Delta \varphi) ] 
\cos^2 ( \varphi / 4N_L ) \Theta( 2\pi N_L - |\varphi| )$, 
where $\varphi = \omega (t+z)$ is the phase of the fundamental light,
fundamental frequency $\omega=\unit{1.55}{\electronvolt}$, 
the relative phase $\Delta \varphi=0$ is chosen for maximum asymmetry and $N_L$ is the number of laser cycles. Note that by this definition, the fraction of the total pulse energy in the second harmonic is $c_2^2/(1+c_2^2)$.  The nonzero %electric and magnetic 
field components are
$B_y = - E_x = E_0 \psi(\varphi)$, with
$\chi_0 = e p_0^+ E_0/m^3 = (p_0^+\omega/m^2) a_0$.
The number of cycles and field strength are related to the pulse duration,   $T_\mathrm{FWHM}[\femto\second] = 2.43 \, N_L \lambda[\micro\metre]$, and laser power  $P[\mathrm{TW}] \simeq 0.0429 \, a_0^2 \, (1+c_2^2) (w_0 /\lambda )^2$. %required to generate such a laser field.

\begin{figure*}[!t]
    \centering
    \parbox{0.66\textwidth}
    {\includegraphics[width=0.66\textwidth]{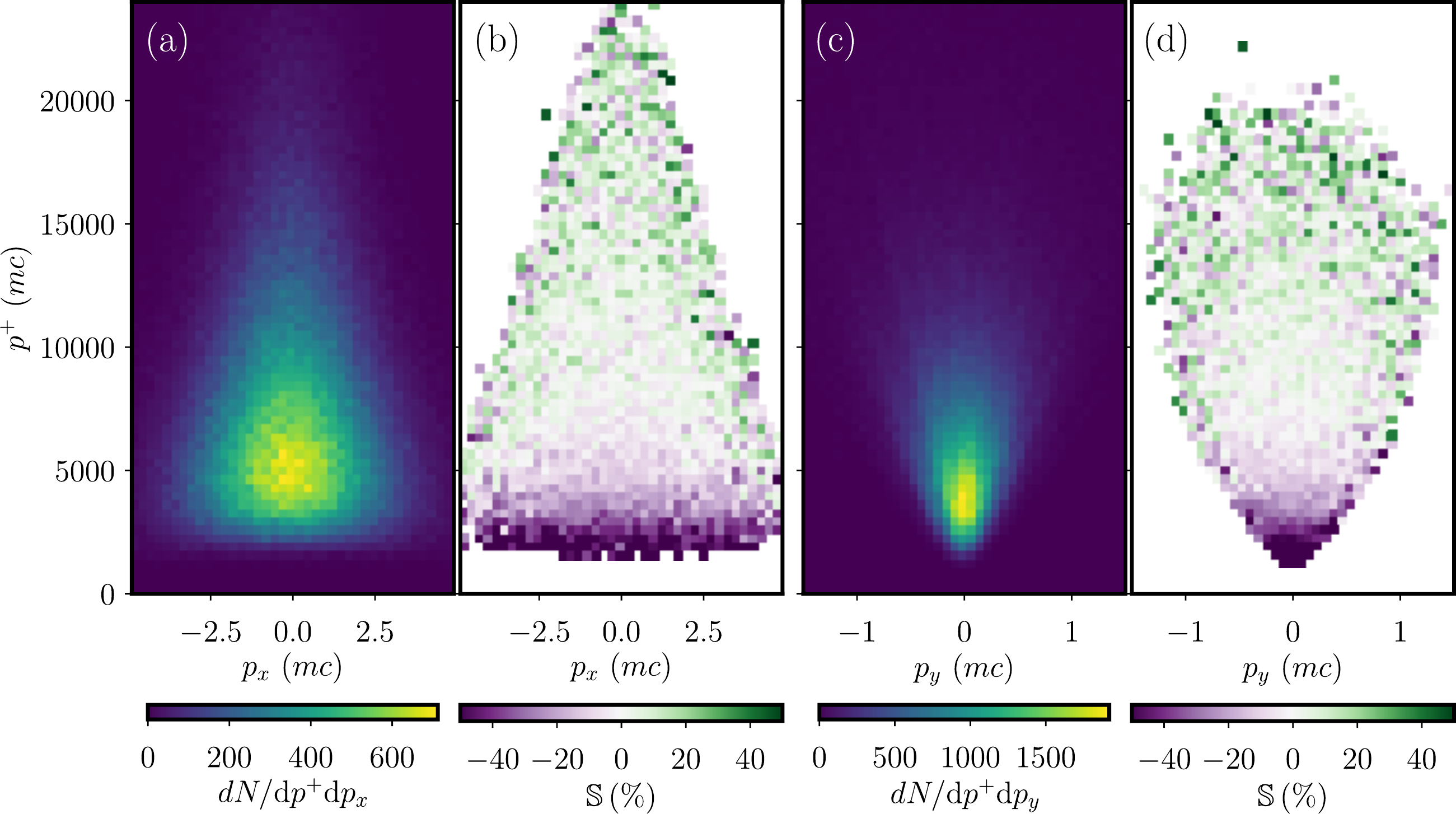} }
    \parbox{0.33\textwidth}
    {
    \includegraphics[width=0.33\textwidth]{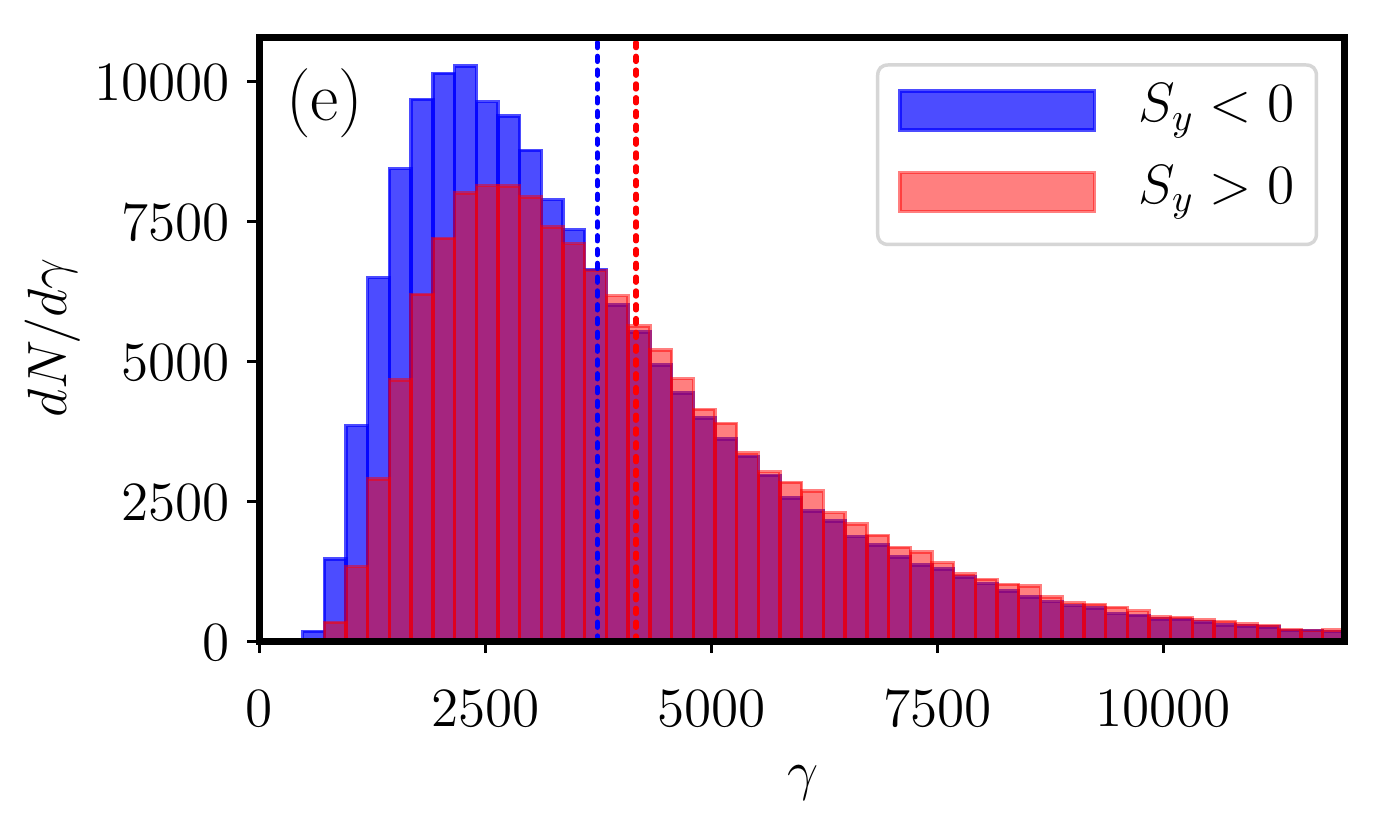}\\
    \includegraphics[width=0.33\textwidth]{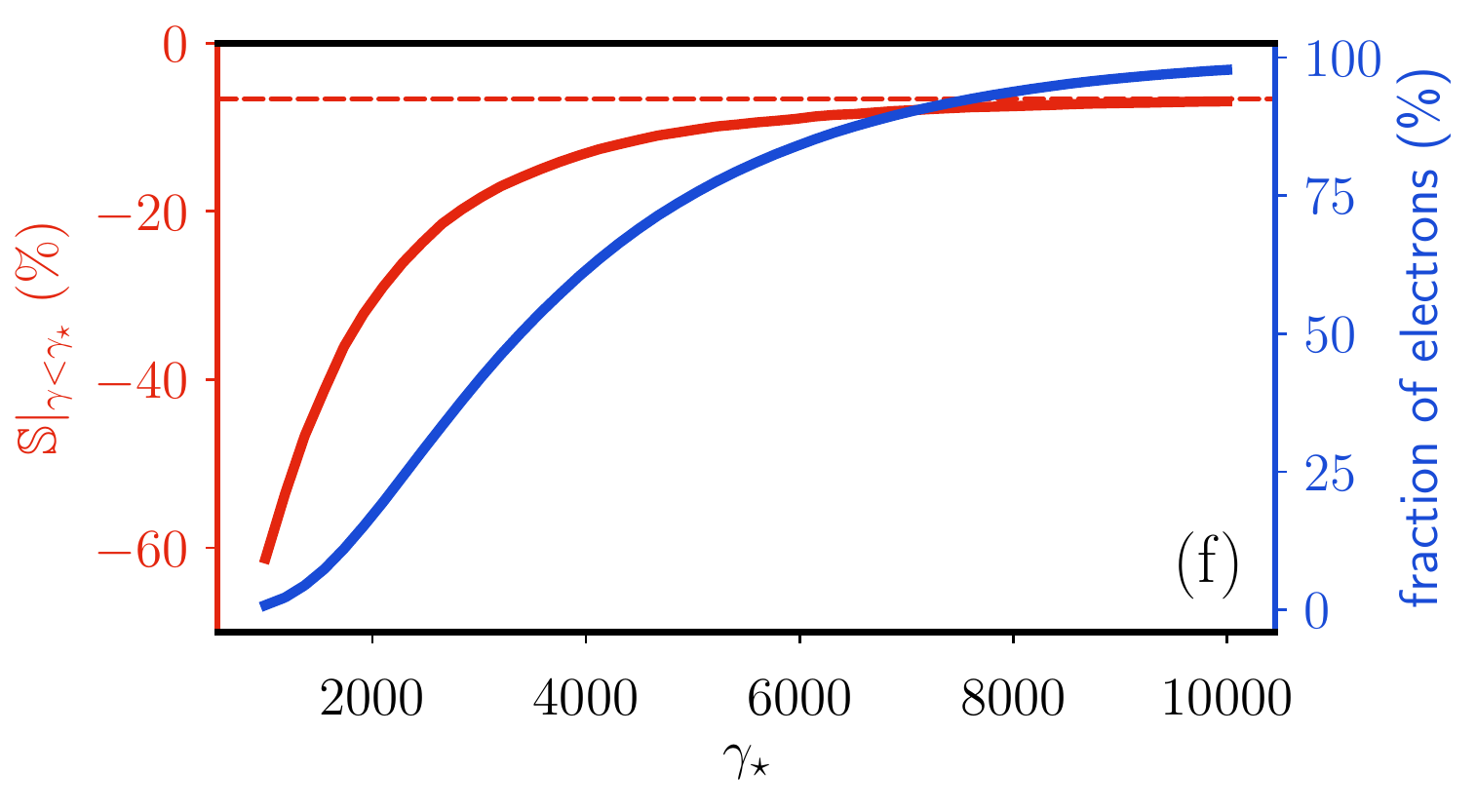}
    }
    \caption{Phase space distributions of scattered electrons (a)-(d) show a correlation of electron momenta and their spin polarization: Low-energy electrons are stronger polarized.
    The final energy distributions for up and down electrons (e) show the effect of spin-dependent radiation reaction: Down electrons lost more energy than spin up electrons on average.
    Lower right: Polarization of post-selected electrons with $\gamma<\gamma_\star$ (left axis) and their relative fraction (right axis).
    }
    \label{fig:hist}
\end{figure*}

Solutions of the kinetic equation as a function of laser phase $\varphi$ are shown in Fig.~\ref{fig:BELLA-KE8} for $n^{+1}(0) = n^{-1}(0)$. 
The distributions $n^{\pm1}$
in (a) and (b) show the energy loss due to quantum radiation reaction effects, with the blue curves representing the mean values. The blue curve in (c) depicts the build-up of the polarization of the
electrons, reaching a final value of $\unit{-7.9}{\%}$.

In Fig.~\ref{fig:survey} we show a parameter scan of the achievable degree of polarization as a function of $\chi_0$ and $c_2$. It shows that $\chi_0>0.5$ is required to acquire some significant polarization for $T=\unit{161}{\femto\second}$. For larger values of $\chi_0 >1.5$ it saturates for $c_2 \simeq 0.7$ at about $\mathbb S\approx \unit{-17}{\%}$. Note that
$\mathbb S\approx {0}$ for a one-color laser pulse ($c_2=0$) for all $\chi_0$.
The survey plot has been calculated for $m\gamma=5$ GeV electrons interacting with a $161$ fs laser pulse. As long as the change of polarization per laser cycle is small, a scaling relation can be found: For fixed $\chi_0$ and $c_2$, the same polarization degree will be achieved for constant values of $T/p^+$, which has been verified numerically.

%\paragraph{Quasi-Classical Monte Carlo Approach---}
In order to describe the electron polarization beyond the 1D plane wave approximation, we extended the accepted quasi-classical model of high-intensity laser matter interactions 
        \cite{Nerush:PRL2011,
        DiPiazza:RevModPhys2012,        
        Ridgers:JCompPhys2014,
        Vranic:NJP2016,
        Harvey:PRA2015,
        Gonoskov:PRE2015,
        King:PRA2013,
        Blackburn:PhysPlas2018} to include the spin degree of freedom.
In this model, electrons follow  classical trajectories between quantum events, at which point emission is treated stochastically using a Monte Carlo algorithm employing the spin-dependent rates, Eq.~\eqref{eq:rates}.
For the classical propagation, we solve the covariant Lorentz force and T-BMT equations using a 4th order Runge-Kutta solver for $u^\mu$ and $S^\mu$, respectively.

Our Monte Carlo algorithm generalizes that presented in   \cite{Duclous:PPCF2011,Ridgers:JCompPhys2014}, to explicitly include the spin of the electrons. It works as follows:
(i) Assign a final optical depth $\tau_\mathrm{em}$ to each particle at the beginning of the simulation and directly after each emission.
(ii) At each timestep after the classical push, project the spin 4-vector onto the local spin quantization direction $s_\zeta = - S.\zeta $ and reduce the remaining optical depth by 
$\sum_{s'} \mathbb R^{s_\zeta s'} \omega u^+ \Delta \tau$.
(iii) A photon is emitted when optical depth reaches zero. Draw a random number $r_1 \in [0,1]$ uniformly. If $r_1 \leq \mathbb R^{s_\zeta,s'=-1}/ \sum_{s'}  \mathbb R^{s_\zeta s'}$
the electron will go to a spin-down state, $s'=-1$, and $s'=+1$ (up) otherwise.
(iv) Sample the normalized energy of the emitted photon from the distribution
$\ud \mathbb R^{s_\zeta s'}/\ud t$ using inverse transform sampling.
(v) Electron recoil is assigned via $\vec p' = (1-t) \vec p $ \cite{Blackburn:PhysPlas2018}, see also \cite{Blackburn:2019},
and the new spin four-vector is determined by $S'_\mu = s' \zeta_\mu'$ where the \emph{new} electron momentum $p'$ is used for calculating $\zeta'$. The Monte Carlo code has been verified against  solutions of the kinetic equation, and predicts the same degree of spin polarization, e.g. $\mathbb S = \unit{-7.9}{\%}$, 
for the case of Fig.~\ref{fig:BELLA-KE8}.

The Monte-Carlo approach allows simulation of realistic three dimensional (3D) scattering scenarios with focused beams. We describe the laser as the superposition of two Gaussian beams in the paraxial approximation for the $1\omega$ and $2\omega$ frequency components with the same focal spot size, $w_0=\unit{8}{\micro \metre}$,
and pulse duration, $T = \unit{109}{\femto\second}$, $a_0=10.8$ and $c_2 = 0.7$, adding up to $0.75$ PW laser power. The two components have different Rayleigh ranges and Guoy phase, which means the $1\omega$ and $2\omega$ components will only be in phase close to the focal plane. The phase shift at the $1\omega$ Rayleigh range is $\Delta \varphi\simeq0.32$. This sets a lower limit on the focal spot size in principle.

We use the Monte Carlo algorithm to simulate the collision of this laser with a 7.8 GeV electron beam with 5 \% energy spread, 0.1 mrad angular divergence and 2.4 $\micro \metre$ beam size, i.e.~$\chi_0=1$ \cite{Bella8GeV}. For scattering from a focused beam,  each electron has $S_y \approx \pm 1$
and it is straightforward to define the two fractions of electrons with spin up and down by $S_y \gtrless 0$. The polarization degree of the beam is calculated as an ensemble average $\mathbb S = \langle  S_y \rangle $.

For this 3D simulation, the final degree of polarization of the whole beam
reaches $\mathbb S = \unit{- 6.6\pm0.4}{\%}$, which is slightly lower than
the prediction for the 1D plane wave case. The statistical uncertainty is due to the finite number of $3\times 10^5$ simulated electrons.
Figure \ref{fig:hist} shows phase spaces $p^+$--$p_x$ (a) and $p^+$--$p_y$ (b)
of the final electrons, as well as the corresponding differential
polarization degrees (b), (d).
For the latter, the degree of polarization is calculated independently for each bin  with at least 20 electrons. These show that the polarization is not uniform over phase space.

Figure \ref{fig:hist} (e) shows the energy spectra for electrons with $S_y \gtrless 0$,
showing that there is a $\unit{6.6}{\%}$ excess of down electrons, yielding the stated polarization degree. Figure \ref{fig:hist} also shows that the mean energy of the down electrons (blue vertical line) is lower than the mean energy of the up electrons (red vertical line) with a relative difference of about $\unit{5.4}{\%}$. The difference in the mean energy of up and down electrons shows that radiation reaction effects are spin dependent.
For $\chi\sim 1$, spin-down electrons radiate about $30\,\%$ more power than the up electrons \cite{book:Sokolov}. 

For an estimate of the spin-dependent radiation reaction, we describe the electron dynamics by the leading term of the quantum corrected Landau-Lifshitz equation 
for each fraction separately \cite{Bulanov:PRE2011,Thomas:PRX2012,Ridgers:JPP2017},
yielding
$\frac{\ud \Delta }{\ud \varphi} \simeq \frac{4\alpha\omega}{3m} a^2(\varphi)
[    ( g_d - g_u ) \bar \gamma^2 -  ( g_d + g_u ) \bar \gamma \Delta  ]$, to linear order in the energy splitting $\Delta = \gamma_u - \gamma_d$,
with spin-dependent Gaunt factors $g_s$ \cite{DelSorbo:PPCF2018}.
The difference in emitted power $\propto g_d-g_u>0$ is partially cancelled by the second term in the square brackets, yielding an equilibrium solution $\Delta  /\bar \gamma~\simeq~( g_d - g_u )/ ( g_d + g_u ) \sim 0.1$ for $\chi\sim1$. 

The phase space distributions in Figure \ref{fig:hist} (a--d) indicate
that the observed degree of polarization can be
increased significantly by post-selecting some electrons, see Figure \ref{fig:hist} (f). If only low-energy electrons are selected with $\gamma<2800$ ($40 \%$ of all electrons), for instance by using a magnetic spectrometer, then their polarization degree is increased to $\mathbb S = - 20 \%$. A similar yet less pronounced enhancement can be achieved by restricting $|p_y|$ to small values.

%\paragraph{Summary/Conclusion---}%
We have shown that it is possible to spin-polarize high-energy electron beams  using intense bi-chromatic laser pulses. We used both a kinetic approach with a linear Boltzmann equation as well as a quasi-classical Monte Carlo approach. 
A parameter scan revealed that a beam polarization of around $\unit{17}{\%}$ can be achieved for $c_2 \simeq 0.7$ and $\chi\gtrsim2$.
We found phase space correlations of the electron polarization which allow to apply
phase space cuts for increasing the measured polarization. We discuss spin-dependent radiation reaction leading to a 5\% mean energy splitting between up and down electrons. In Ref.~\cite{Chen2019}, the generation of positrons with polarization degrees of up to \unit{60}{\%} was proposed by exploiting the spin-asymmetry in the strong-field pair-production process.

The proposed electron beam polarization could be realized experimentally with present day technology in an all-optical set-up: Multi-GeV electron beams (up to 8 GeV) have been demonstrated using laser wakefield accelerators \cite{Bella8GeV}. 
Similarly, one could realize this experiment with a PW laser at a conventional accelerator facility, for instance at SLAC or XFEL \cite{LUXE}. 
The petawatt class laser pulse required needs to be frequency doubled, and the required technology for generating the $2\omega$ light is  second harmonic generation in a nonlinear crystal (having typically $\unit{50}{\%}$ energy efficiency). Growth of crystals large enough to be compatible with PW lasers has been developed \cite{PW_BBO}.
The polarization of the GeV electron beam could be measured using (nonlinear) Compton scattering \cite{Mane:RPP2005,Narayan:PRX2016,Li:2019b},
with \cite{Li:2019b} anticipating an accuracy of \unit{0.3}{\%} from multi-GeV electrons in a single shot.

{As this manuscript was being prepared, Ref.~\cite{Chen2019}
appeared on the arXiv, the subject of which is similar to our own.
A.~G.~R.~T.~acknowledges useful discussions with S. Meuren. 
This work supported by the U.S. Army Research Office Grant No. W911NF-16-1-0044 (DS,AGRT),
by the DOE, LDRD program at SLAC, under contract DE-AC02-76SF00515 (DDS), 
and by the UK Engineering and Physical Sciences Research Council Grant No. EP/M018156/1 (CPR).
}

%\bibliography{references}
%\input{references}

%merlin.mbs apsrev4-1.bst 2010-07-25 4.21a (PWD, AO, DPC) hacked
%Control: key (0)
%Control: author (0) dotless jnrlst
%Control: editor formatted (1) identically to author
%Control: production of article title (0) allowed
%Control: page (1) range
%Control: year (0) verbatim
%Control: production of eprint (0) enabled
%

\end{document}